\documentclass[conference]{IEEEtran}
\usepackage[T1]{fontenc}

\usepackage{amsmath,amssymb}
\usepackage{graphicx}
\usepackage{booktabs}
\usepackage{subcaption}
\usepackage{hyperref}
\usepackage{cite}
\usepackage{listings}
\usepackage{xcolor}
\usepackage{tikz}
\usetikzlibrary{positioning, fit, backgrounds, arrows.meta, calc}
\usepackage{footnote}
\makesavenoteenv{tabular}
\makesavenoteenv{table}

\graphicspath{{images/}}

\title{A Modular Open-Source Python Framework for Small Satellite Attitude Determination and Control}
\author{
\IEEEauthorblockN{Patrick McKeen}
\IEEEauthorblockA{MIT AeroAstro\\pmckeen@mit.edu}
\and
\IEEEauthorblockN{Niclas Scheuer}
\IEEEauthorblockA{ETH Z\"{u}rich D-MAVT\\nscheuer@ethz.ch}
\and
\IEEEauthorblockN{Kerri Cahoy}
\IEEEauthorblockA{MIT AeroAstro\\kcahoy@mit.edu}
}

\begin{document}
\maketitle
\vspace{-5em}

\begin{abstract}
Advanced attitude determination and control (ADCS) techniques are often difficult for small satellite teams to adopt due to tightly coupled implementations and the need to rapidly evaluate alternative architectures. Evaluating prospective ADCS options may require significant modification of existing code for specific actuator combinations, sensor hardware, orbits, mission goals, and control laws. Existing tools have broader scopes: Basilisk uses a compiled C/C++ core with Python interfaces, while OpenSatKit integrates NASA's core Flight System, COSMOS, and the 42 simulator. Their architectures support general mission simulation or flight software development, but can require substantial setup for closed-loop ADCS studies. This paper presents an open-source, modular ADCS framework in pure Python that enables rapid development, repeatable testing, and flight-oriented integration across heterogeneous spacecraft.

The package supports torque- and command-level control, bound-respecting allocation, and full- and reduced-attitude objectives. Estimation, control, planning, and actuator/sensor modeling are implemented as interchangeable modules reconfigured through configuration rather than code restructuring. Integrated simulation includes attitude and orbit propagation, configurable sensor and actuator models with bias and noise, and environmental disturbances. An actuator-aware planner supports underactuated architectures, while hardware-in-the-loop capability enables testing on mission-representative computers.

We demonstrate the framework through a variety of case studies including a 3U CubeSat with three magnetorquers and one reaction wheel, a magnetorquer-only 1U, and a 6U with three magnetorquers and three reaction wheels. Across these, we showcase actuator and sensor failure modes, full- and reduced-attitude goals, and momentum management. HIL testing on a Raspberry Pi is demonstrated in the same package. Adapting a published control law to a new actuator configuration requires less than 5 lines of configuration changes. The framework underpins flight software for an Earth-observing CubeSat with visible and long-wave infrared imagers, three magnetorquers, and one reaction wheel, without star trackers or propulsion.
\end{abstract}

\section{Introduction}\label{sec:intro}
Many satellites need to control their attitude to meet their mission goals, whether that means pointing an imager at a target (and not at the sun!), a radio at a ground station, or solar panels towards the sun. While critical for most missions, ADCS is considered a difficult and arcane area of work. Development of a working ADCS can be painful, especially for resource-constrained teams. Even with the wealth of existing research and many control options available in literature, differences in experimental setup means that evaluating alternative architectures means reimplementing, not just reconfiguring. Many published control laws and other ADCS components assume specific hardware, orbits, goals, and more---translating that research to your spacecraft's particulars can take days or weeks of custom development.

This leads to an issue for many small satellite teams, where they cannot explore the full option space---testing every idea they find in literature or flight heritage is too expensive in terms of time and energy, so they often pick something conservative and never explore alternatives. Existing tools for satellite development tend to not specifically focus on ADCS. Even those with such capability (Basilisk~\cite{basilisk}) are powerful but heavyweight (C++ cores, complex message-passing, etc.), resulting in steep learning curves for ADCS-focused tasks.

This paper presents a pure-Python modular ADCS framework designed to make ADCS evaluation a question of configuration and testing rather than repeated implementation. The framework contains interchangeable components for estimation, control, planning, and simulation (including hardware-in-the-loop), and is accessible enough to go from \texttt{git clone} to a working simulation in less than fifteen minutes. With fast and cheap iteration, teams can explore more of the design space---and find better solutions they never would have tried.

\section{Framework Overview}\label{sec:overview}

\subsection{Architecture}\label{subsec:architecture}

The framework is written in pure Python, with no compiled C/C++ dependencies beyond NumPy and SciPy. This is a deliberate design choice: every line of the ADCS stack is readable, debuggable, and modifiable. Users can set breakpoints inside the estimator, inspect intermediate torque allocations, or add print statements to the dynamics---things that are difficult or impossible when the core logic lives behind a compiled interface. The tradeoff is speed, but as we show in Section~\ref{subsec:demo_hil}, Python is fast enough for CubeSat-class control loops on flight-representative hardware.

The framework is organized around a simulation loop that executes the same sequence at each timestep: propagate the orbit, read sensors, estimate the state, evaluate the current goal, compute control commands, and integrate the dynamics forward. Each of these stages is represented by a modular, interchangeable Python object. Replacing any component---a different estimator, a different control law, a different actuator set---requires changing the object, not restructuring the code.

The central abstraction is the separation between the \emph{true} spacecraft and the \emph{estimated} spacecraft. The \texttt{Satellite} object holds the ground-truth physical model: mass properties, actuators with their true limits and noise characteristics, sensors with their true biases, and disturbance models. The \texttt{EstimatedSatellite} represents what the onboard software believes about the spacecraft---the same structure, but with estimated or pre-assigned (and possibly incorrect) parameters and optional bias states that the estimator can track. Controllers and estimators operate on the estimated satellite; the simulation environment operates on the true one. This split is fundamental to realistic testing: it means that sensor noise (which can differ between the true and estimated satellite), actuator misalignment, inertia errors, and unmodeled disturbances all affect the control loop naturally, without special injection code. A simulation with perfect knowledge is simply one where the estimated satellite matches the truth.

The simulation function itself is compact:
\begin{lstlisting}
results = ADCS.simulate(
    x=x_0,              # initial state
    satellite=real_sat,  # true spacecraft model
    controller=ctrl,     # control law (any Controller subclass)
    estimator=est,       # state estimator (optional)
    goal=goal,           # pointing objective (or GoalList)
    os0=os0,             # initial orbital state
    dt=1.0, tf=1000.0    # timestep and duration
)
\end{lstlisting}
Every argument after the initial state and true satellite is optional. Omitting the estimator runs the controller on the true state (perfect knowledge). Omitting the controller runs an open-loop simulation. This progressive complexity---from open-loop dynamics through perfect-knowledge control to full estimation-in-the-loop---allows users to build understanding incrementally and isolate the effect of each subsystem.

The control law interface is a single method: given the estimated state, sensor measurements, estimated satellite model, orbital state, and current goal, return actuator commands. All implemented controllers---PD, magnetic (Lovera~\cite{lovera2004periodic}, Wisniewski~\cite{wisniewski1998sliding}), B-dot detumbling, LP/QP allocation variants, the ALTRO trajectory planner~\cite{howell_altro_2019}, and more---conform to this interface. Swapping between them is a one-line change in the simulation setup. The allocation layer (LP, QP, or constrained variants) is easily embedded within each controller that requires it, so the user selects a controller and gets the full pipeline from desired torque to actuator commands.

Goals are similarly modular. The framework supports full-attitude goals (target quaternion), reduced-attitude goals (align a body axis with an inertial direction), and a library of common pointing objectives: nadir, sun, anti-sun, magnetic field, ground coordinates, and arbitrary ECI directions. A \texttt{GoalList} allows time-varying goal sequences---for example, tracking a ground target with a camera during an imaging pass and then reorienting solar panels toward the sun:

\begin{lstlisting}
goal_timeline = {
    0.0:   ADCS.goals.Coordinate_Goal(lat=33.75, lon=-84.39),
    500.0: ADCS.goals.Sun_Goal()
}
goal = ADCS.GoalList(goal_timeline, time_units="seconds")
\end{lstlisting}

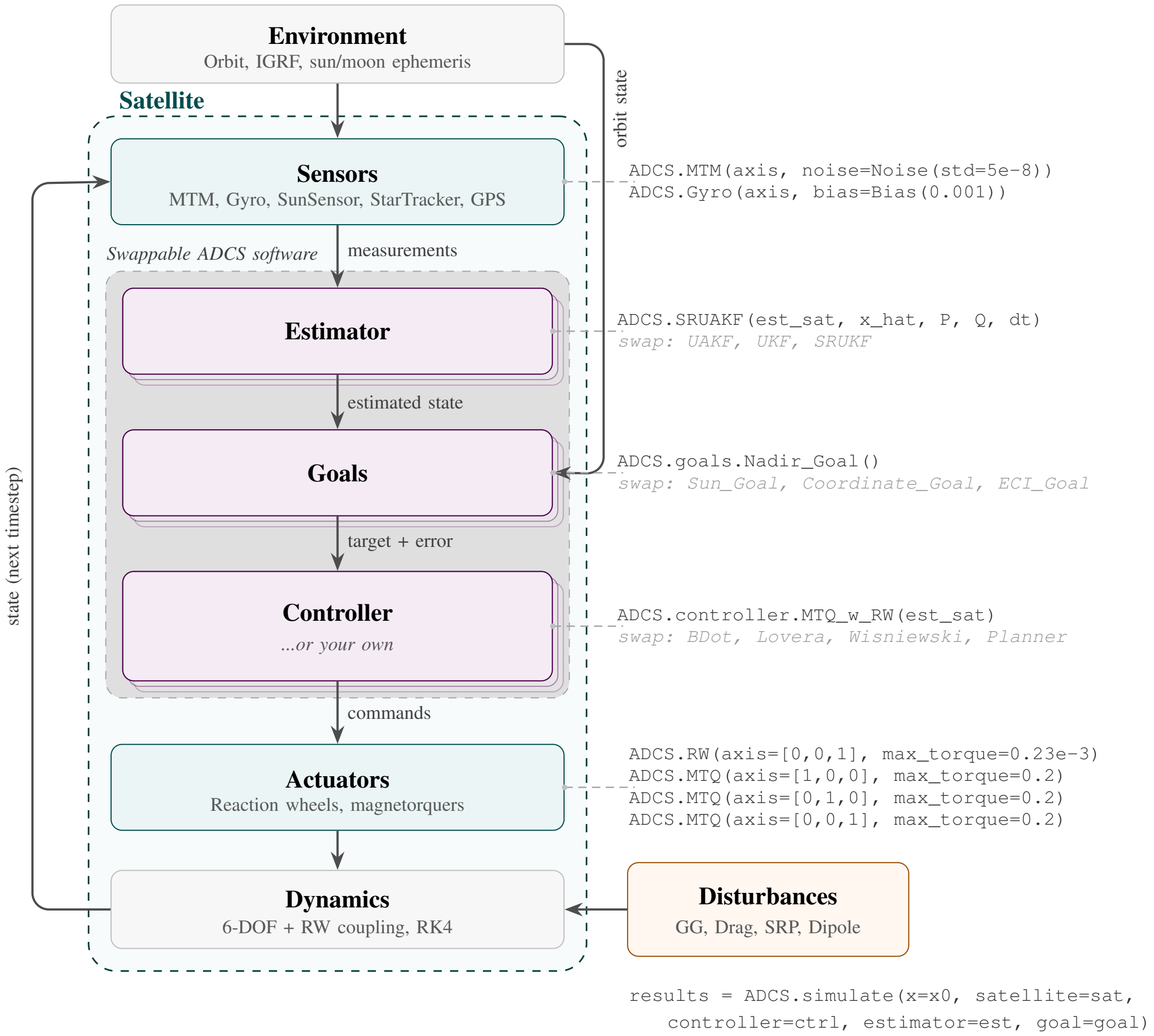
\begin{figure*}[t]
\centering
\begin{tikzpicture}[
    >=Stealth,
    node distance=0.5cm,
    hw/.style={draw=teal!60!black, fill=teal!8, rounded corners=4pt,
               minimum width=5.8cm, minimum height=1.1cm, line width=0.4pt,
               font=\small},
    sw/.style={draw=violet!60!black, fill=violet!8, rounded corners=4pt,
               minimum width=5.5cm, minimum height=1.1cm, line width=0.4pt,
               font=\small},
    sw tall/.style={sw, minimum height=1.4cm},
    env/.style={draw=gray!50, fill=gray!6, rounded corners=4pt,
                minimum width=5.8cm, minimum height=1.0cm, line width=0.4pt,
                font=\small},
    dist/.style={draw=orange!60!black, fill=orange!6, rounded corners=4pt,
                 minimum width=3.6cm, minimum height=1.2cm, line width=0.4pt,
                 font=\small},
    boxtitle/.style={font=\small\bfseries},
    boxsub/.style={font=\scriptsize, text=gray!70!black},
    code/.style={font=\scriptsize\ttfamily, text=gray!40!black, anchor=west},
    codealt/.style={font=\scriptsize\ttfamily\itshape, text=gray!80!white, anchor=west},
    flow/.style={->, gray!60!black, line width=0.8pt},
    flowlabel/.style={font=\scriptsize, text=gray!50!black},
    leader/.style={gray!70, line width=0.5pt, densely dashed},
    loopback/.style={->, gray!60!black, line width=0.8pt},
]

\node[env] (environment) at (0, 0) {};
\node[boxtitle] at ([yshift=3pt]environment.center) {Environment};
\node[boxsub] at ([yshift=-7pt]environment.center) {Orbit, IGRF, sun/moon ephemeris};
\node[boxtitle] at ([yshift=4pt]environment.center) {};

\node[hw, below=0.7cm of environment] (sensors) {};
\node[boxtitle] at ([yshift=3pt]sensors.center) {Sensors};
\node[boxsub] at ([yshift=-7pt]sensors.center) {MTM, Gyro, SunSensor, StarTracker, GPS};

\node[sw, below=0.8cm of sensors] (est shadow2) {};
\begin{scope}[opacity=0.25]
\node[sw] at ([xshift=4pt, yshift=-4pt]est shadow2.center) {};
\end{scope}
\begin{scope}[opacity=0.4]
\node[sw] at ([xshift=2pt, yshift=-2pt]est shadow2.center) {};
\end{scope}
\node[sw] (estimator) at (est shadow2.center) {};
\node[boxtitle] at ([yshift=0pt]estimator.center) {Estimator};
\node[boxsub] at ([yshift=-7pt]estimator.center) {};

\node[sw, below=0.7cm of estimator] (goal shadow) {};
\begin{scope}[opacity=0.25]
\node[sw] at ([xshift=4pt, yshift=-4pt]goal shadow.center) {};
\end{scope}
\begin{scope}[opacity=0.4]
\node[sw] at ([xshift=2pt, yshift=-2pt]goal shadow.center) {};
\end{scope}
\node[sw] (goals) at (goal shadow.center) {};
\node[boxtitle] at ([yshift=0pt]goals.center) {Goals};
\node[boxsub] at ([yshift=-7pt]goals.center) {};

\node[sw tall, below=0.7cm of goals] (ctrl shadow) {};
\begin{scope}[opacity=0.25]
\node[sw tall] at ([xshift=4pt, yshift=-4pt]ctrl shadow.center) {};
\end{scope}
\begin{scope}[opacity=0.4]
\node[sw tall] at ([xshift=2pt, yshift=-2pt]ctrl shadow.center) {};
\end{scope}
\node[sw tall] (controller) at (ctrl shadow.center) {};
\node[boxtitle] at ([yshift=5pt]controller.center) {Controller};
\node[boxsub, font=\scriptsize\itshape] at ([yshift=-8pt]controller.center) {...or your own};

\node[hw, below=0.8cm of controller, minimum height=1.1cm] (actuators) {};
\node[boxtitle] at ([yshift=3pt]actuators.center) {Actuators};
\node[boxsub] at ([yshift=-7pt]actuators.center) {Reaction wheels, magnetorquers};

\node[env, below=0.5cm of actuators] (dynamics) {};
\node[boxtitle] at ([yshift=3pt]dynamics.center) {Dynamics};
\node[boxsub] at ([yshift=-7pt]dynamics.center) {6-DOF + RW coupling, RK4};

\node[dist, right=0.8cm of dynamics, yshift=0cm] (disturbances) {};
\node[boxtitle] at ([yshift=5pt]disturbances.center) {Disturbances};
\node[boxsub] at ([yshift=-7pt]disturbances.center) {GG, Drag, SRP, Dipole};

\begin{scope}[on background layer]
\node[draw=teal!50!black, dashed, line width=0.6pt, rounded corners=10pt,
      fill=teal!3, inner sep=8pt,
      fit=(sensors)(actuators)(dynamics),
      ] (satbox) {};
\node[font=\small\bfseries, text=teal!60!black, anchor=south west]
    at ([xshift=8pt, yshift=-1pt]satbox.north west) {Satellite};
\end{scope}

\begin{scope}[on background layer]
\node[draw=gray!70, dashed, line width=0.4pt, rounded corners=6pt,
      fill=gray!25, inner sep=6pt,
      fit=(estimator)(goals)(controller)(est shadow2)(ctrl shadow),
      ] (swbox) {};
\node[font=\scriptsize\itshape, text=gray!50!black, anchor=south west]
    at ([xshift=-3pt, yshift=-1pt]swbox.north west) {Swappable ADCS software};
\end{scope}

\draw[flow] (environment) -- (sensors);
\draw[flow] (sensors) -- node[right, flowlabel,yshift=2pt] {measurements} (estimator);
\draw[flow] (estimator) -- node[right, flowlabel] {estimated state} (goals);
\draw[flow] (goals) -- node[right, flowlabel] {target + error} (controller);
\draw[flow] (controller) -- node[right, flowlabel] {commands} (actuators);
\draw[flow] (actuators) -- (dynamics);
\draw[flow] (disturbances) -- (dynamics);

\draw[loopback, rounded corners=6pt]
    (dynamics.west) -- ++(-1.0, 0) coordinate (loop-bot) -- ++(0, 9.3) coordinate (loop-top) -- (sensors.west);
\node[flowlabel, rotate=90, anchor=south]
    at ($(loop-bot)!0.5!(loop-top)$) {state (next timestep)};

\draw[flow, rounded corners=6pt]
    (environment.east) -- ++(0.5, 0) coordinate (orb-bot) -- ++(0, -5.5) coordinate (orb-top) -- (goals.east);
\node[flowlabel, rotate=90, anchor=north]
    at ($(orb-bot)!0.15!(orb-top)$){orbit state};

\draw[leader] (sensors.east) -- ++(0.9, 0) coordinate (s1);
\fill[gray!50] (sensors.east) circle (1pt);
\node[code] at ([xshift=0.7cm, yshift=4pt]sensors.east) {ADCS.MTM(axis, noise=Noise(std=5e-8))};
\node[code] at ([xshift=0.7cm, yshift=-4pt]sensors.east) {ADCS.Gyro(axis, bias=Bias(0.001))};

\draw[leader] (estimator.east) -- ++(0.9, 0);
\fill[gray!50] (estimator.east) circle (1pt);
\node[code] at ([xshift=0.7cm,yshift=4pt]estimator.east) {ADCS.SRUAKF(est\_sat, x\_hat, P, Q, dt)};
\node[codealt] at ([xshift=0.7cm, yshift=-4pt]estimator.east) {swap: UAKF, UKF, SRUKF};

\draw[leader] (goals.east) -- ++(0.9, 0);
\fill[gray!50] (goals.east) circle (1pt);
\node[code] at ([xshift=0.7cm,yshift=4pt]goals.east) {ADCS.goals.Nadir\_Goal()};
\node[codealt] at ([xshift=0.7cm, yshift=-4pt]goals.east) {swap: Sun\_Goal, Coordinate\_Goal, ECI\_Goal};

\draw[leader] (controller.east) -- ++(0.9, 0);
\fill[gray!50] (controller.east) circle (1pt);
\node[code] at ([xshift=0.7cm,yshift=4pt]controller.east) {ADCS.controller.MTQ\_w\_RW(est\_sat)};
\node[codealt] at ([xshift=0.7cm, yshift=-4pt]controller.east) {swap: BDot, Lovera, Wisniewski, Planner};

\draw[leader] (actuators.east) -- ++(0.9, 0);
\fill[gray!50] (actuators.east) circle (1pt);
\node[code] at ([xshift=0.7cm,yshift=12pt]actuators.east) {ADCS.RW(axis=[0,0,1], max\_torque=0.23e-3)};
\node[code] at ([xshift=0.7cm, yshift=4pt]actuators.east) {ADCS.MTQ(axis=[1,0,0], max\_torque=0.2)};
\node[code] at ([xshift=0.7cm, yshift=-4pt]actuators.east) {ADCS.MTQ(axis=[0,1,0], max\_torque=0.2)};
\node[code] at ([xshift=0.7cm, yshift=-12pt]actuators.east) {ADCS.MTQ(axis=[0,0,1], max\_torque=0.2)};

\node[code] at ([xshift=0.7cm, yshift=-0.6cm]dynamics.east |- dynamics.south)
    {results = ADCS.simulate(x=x0, satellite=sat,};
\node[code] at ([xshift=0.7cm, yshift=-0.95cm]dynamics.east |- dynamics.south)
    {\quad\quad controller=ctrl, estimator=est, goal=goal)};

\end{tikzpicture}
\caption{Framework architecture. Each purple box is a swappable module configured with one line of Python. Stacked outlines indicate interchangeable components. Code callouts show the actual API for each stage.}
\label{fig:architecture}
\end{figure*}

\subsection{What's Included}\label{subsec:included}

The framework provides a complete ADCS development environment, summarized in Table~\ref{tab:components}.

\begin{table}[h]
\centering
\caption{Framework components.}
\label{tab:components}
\begin{tabular}{p{1.6cm} p{6.0cm}}
\toprule
Component & Capabilities \\
\midrule
Dynamics & 6-DOF attitude with RW coupling, J2 orbit propagation \\
Sensors & Magnetometer, gyro, sun sensor, star tracker, Earth horizon sensors, GPS---configurable bias, noise, and mounting geometry \\
Actuators & Reaction wheels and magnetorquers with saturation limits, noise, and arbitrary axis placement \\
Disturbances & Gravity gradient, drag, SRP, residual dipole, propulsion, generalized and estimated disturbances---individually toggleable \\
Estimation & UKF, SRUKF, and augmented-state variants for joint attitude, bias, and disturbance tracking \\
Control & PD, Lovera, Wisniewski, B-dot, hybrid MTQ-RW, ALTRO trajectory planner \\
Allocation & LP, QP, weighted and constrained QP---bound-respecting torque allocation \\
Goals & Nadir, sun, ground track, ECI, reduced-attitude pointing---time-varying sequences via \texttt{GoalList}, as well as keep-out zones \\
Infrastructure & Satellite factory with COTS models, Monte Carlo interface, HIL on Raspberry Pi \\
Monte Carlo & Dedicated \texttt{MCConfig} interface with random ICs, orbits, and goals\\
\bottomrule
\end{tabular}
\end{table}

\subsection{Hardware-in-the-Loop Capability}\label{subsec:hil_overview}

Because the framework is pure Python, the same code that runs in desktop simulation runs unmodified on flight hardware. The BeaverCube~2 mission uses this framework as its flight ADCS software, executing on a Raspberry Pi. In hardware-in-the-loop (HIL) testing, an external computer simulates the physical environment (dynamics, sensor outputs, actuator effects) while the flight computer runs the complete ADCS stack---estimation, control, and allocation---receiving simulated sensor data and commanding simulated actuators over a serial link. This eliminates the ``simulation vs.\ flight'' code divergence that plagues many missions and ensures that the software tested on the ground is the software that flies. The HIL infrastructure, shown in Figure~\ref{fig:hil}, is included in the open-source package, so any team with a single-board computer and a serial connection can replicate this testing setup.

\section{Getting Started: 5 Minutes to a Working Simulation}\label{sec:quickstart}

After installation (\texttt{pip install -e .}), a complete closed-loop ADCS simulation requires fewer than 20 lines of Python. The following example sets up a 3U CubeSat with three magnetorquers and one reaction wheel, points it at a ground target, and runs for 500 seconds:

\begin{lstlisting}
import ADCS
import numpy as np
 
# Actuators: 3 magnetorquers + 1 reaction wheel
acts = [ADCS.MTQ(axis, max_torque=0.2) for axis in np.eye(3)]
acts += [ADCS.RW(axis=np.array([0,0,1]), max_torque=0.23e-3,
                  J=5.7e-6, h_max=3.5e-3)]
 
# Sensors: 3-axis magnetometer
sens = [ADCS.MTM(axis) for axis in np.eye(3)]
 
# Spacecraft
sat = ADCS.Satellite(mass=4,
        J_0=np.diag([0.03, 0.03, 0.01]),
        actuators=acts, sensors=sens,
        boresight=np.array([0, 0, 1]))
 
# Initial state: [angular velocity, quaternion, wheel momentum]
x_0 = np.array([0.01, -0.02, 0.01, 1, 0, 0, 0, 0.0])
 
# Controller, goal, orbit
ctrl = ADCS.controller.MTQ_w_RW_LP(
        est_sat=sat, p_gain=5e-5, d_gain=2e-3, c_gain=1e-3)
goal = ADCS.goals.Coordinate_Goal(lat=42.36, lon=-71.06)
os0 = ADCS.Orbital_State(ephem=ADCS.Ephemeris(), J2000=0.22,
        R=np.array([5000, 0, 5000]), V=np.array([0, 7.5, 0]))
 
# Run
results = ADCS.simulate(x=x_0, satellite=sat,
             controller=ctrl, goal=goal, os0=os0,
             dt=1.0, tf=500.0)
\end{lstlisting}

\begin{figure*}[]
\centering
\includegraphics[clip, trim={0 0 12.25cm 0},width=7in]{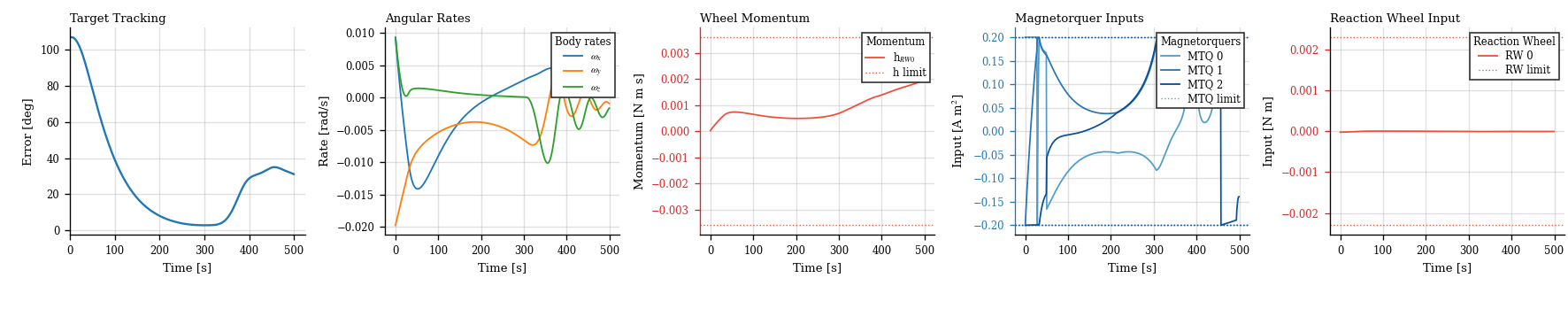}
\caption{Output plots from the quickstart example. It shows how quickly one can get readable (though not yet optimal) results.}
\label{fig:quickstart}
\end{figure*}

This produces a \texttt{SimulationResults} object containing time histories of state, control commands, and pointing error. The built-in plotting interface provides standard diagnostic views, with results seen in Figure~\ref{fig:quickstart}:

\begin{lstlisting}
ADCS.plot(results,
    ADCS.plots.TargetPlot(),
    ADCS.plots.AngularVelocityPlotCombined(sources=["real"]),
    ADCS.plots.ControlPlotCombined(),
    layout=(1, 3))
\end{lstlisting}

Everything from here is a configuration change. Swapping to a magnetorquer-only system means removing the \texttt{RW} line. Swapping to a different control law means replacing the \texttt{controller} assignment. Adding estimation means constructing an \texttt{EstimatedSatellite} and \texttt{SRUAKF} and passing them to \texttt{simulate}---the same function call, two additional arguments. Pre-configured spacecraft are also available via the satellite factory: \texttt{create\_beavercube2\_cubesat()} returns a complete 3+1 spacecraft with realistic COTS hardware parameters (ISIS magnetorquer board, CubeSpace CubeWheel, ICM-20948 IMU).

\section{Demonstrations}\label{sec:demos}

\subsection{Rapid Architecture Trade}\label{subsec:demo_arch_trade}

A common early-mission question is what ADCS actuation architecture to use---are reaction wheels required and, if so, how many? Magnetorquer-only control (3+0) minimizes cost and complexity; adding one wheel (3+1) or three (3+3) buys pointing performance but adds mass, power, and failure modes. Answering this question traditionally requires building separate simulations for each configuration. In this framework, the entire trade is a change to the actuator list:

\begin{lstlisting}
# 3+0: magnetorquers only
actuators = mtqs

# 3+1: add one reaction wheel on the z-axis
actuators = mtqs + [RW(axis=[0,0,1], max_torque=0.23e-3,
                       max_momentum=3.5e-3)]

# 3+3: three orthogonal reaction wheels
actuators = mtqs + [RW(axis=[1,0,0], ...),
                    RW(axis=[0,1,0], ...),
                    RW(axis=[0,0,1], ...)]
\end{lstlisting}

Everything else---the controller, estimator, disturbances, orbit, goals, and simulation parameters---remains identical. The framework's generalized allocation layer automatically adapts to whatever actuators are present, so no changes to the control law are needed. The LP allocation may not produce the optimal controller for any single configuration --- a hand-tuned law exploiting specific actuator geometry and focused on mission goals could easily outperform it. But it produces reasonable performance across all configurations without per-mission, per-actuator, or per-goal tuning, which is the point during the design phase: to identify which architectures merit further investment, not to extract maximum performance from each one on the first pass.

\begin{figure}[]
\centering
\includegraphics[width=\linewidth]{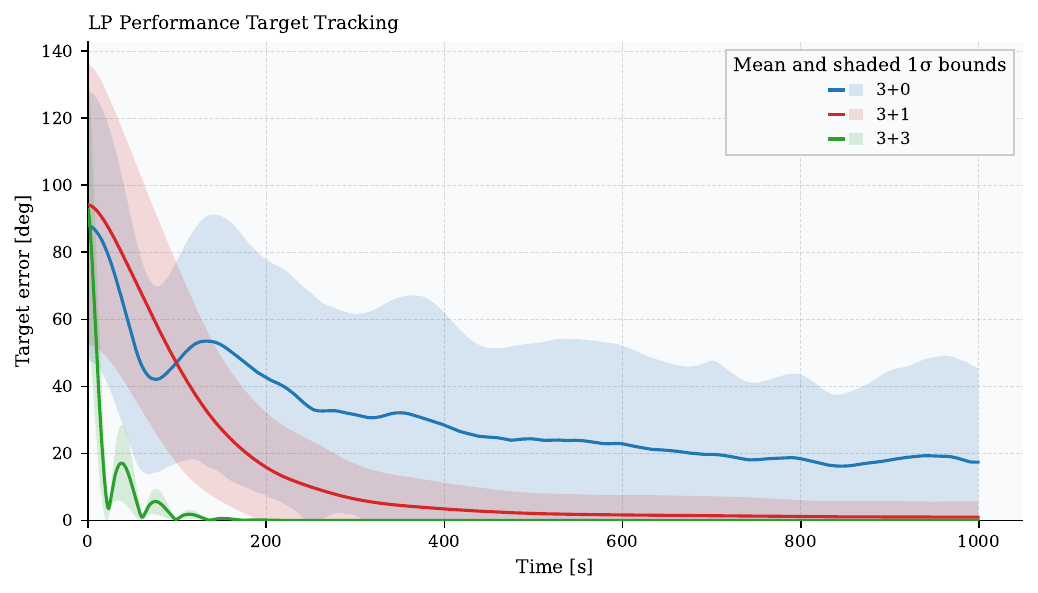}
\caption{Side-by-side MC pointing error distributions for 3+0, 3+1, 3+3.}
\label{fig:trade}
\end{figure}

To quantify the difference, we can run 100-trial Monte Carlo comparisons with randomized initial conditions and pointing goals over a 1000-second horizon (roughly one-sixth of a LEO orbit). All configurations use the same LP-allocated PD controller. The pointing errors of the respective architectures can be found in Figure~\ref{fig:trade}, with the results summarized in Table~\ref{tab:arch_trade}.

\begin{table}[h]
\centering
\caption{Pointing performance across actuator configurations (100-trial Monte Carlo, 1000\,s, PD control with LP allocation).}
\label{tab:arch_trade}
\begin{tabular}{lccc}
\hline
Configuration & Mean Error (deg) & \% $<$ 1$^\circ$ & \% $<$ 5$^\circ$ \\
\hline
3+0 (MTQ-only) & 17.4 & 19 & 42 \\
3+1 (MTQ + 1 RW) & 1.0 & 95 & 98 \\
3+3 (MTQ + 3 RW) & 0.067 & 100 & 100 \\
\hline
\end{tabular}
\end{table}

Adding a single reaction wheel reduces mean pointing error by more than an order of magnitude and brings the majority of trials within one degree. Adding two more wheels improves further, but the marginal gain is smaller: the jump from zero wheels to one is the dominant improvement. This reflects a qualitative change in instantaneous torque authority. Magnetorquers generate torque only perpendicular to the magnetic field, while the reaction wheel supplies torque along its fixed axis. Together they span all three torque directions almost everywhere in magnetic-field direction space: for nonzero $\mathbf{B}$, the combined actuation map has full rank whenever $\hat{\mathbf{a}}_{\mathrm{RW}}\cdot\mathbf{B}\ne 0$, where $\hat{\mathbf{a}}_{\mathrm{RW}}$ is the wheel axis. The perpendicular configurations form a measure-zero set; available torque remains limited by actuator bounds and wheel momentum capacity. The total code change to discover this is three lines. The total wall-clock time, including all 300 simulation runs, is less than an hour on a laptop.

\subsection{Control Law Comparison}\label{subsec:demo_control_law}
Swapping control laws is a one-line configuration change: the same 3+1 hardware, allocation, compensation, estimation, and disturbance environment can run a Wie quaternion PD~\cite{wie2008spacecraft}, Lovera magnetic controller, or Wisniewski sliding-mode controller with no other modifications. This enables fair, apples-to-apples comparison between literature controllers---an evaluation that previously required reimplementing each law for a specific actuator set.

\subsection{Realistic Testing Toggle}\label{subsec:demo_realism}

Most ADCS development begins---and sometimes ends---with idealized simulation: perfect state knowledge, no disturbances, no sensor noise. The framework makes it straightforward to ratchet up fidelity incrementally, so users can isolate the effect of each source of realism. The progression requires no structural changes to the simulation:

\begin{lstlisting}
# Level 1: Perfect knowledge, no disturbances
results = ADCS.simulate(x=x_0, satellite=sat,
             controller=ctrl, goal=goal, os0=os0)
 
# Level 2: Add sensor noise
sat = ADCS.Satellite(..., sensors=noisy_sens)
 
# Level 3: Add estimation (controller sees estimated state)
est = ADCS.SRUAKF(est_sat=est_sat, x_hat=x_hat,
                   P_hat=P0, Q_hat=Q, dt=dt)
 
# Level 4: Add disturbances
sat = ADCS.Satellite(..., disturbances=dists)
\end{lstlisting}

The gap between Level~1 and Level~4 performance---often surprisingly large---represents risk that goes unquantified when teams test only under idealized conditions. Making it trivial to run all four levels means teams can isolate the effect of each degradation and achieve higher confidence in less time.





\subsection{Multi-Scale Configurations}\label{subsec:demo_multiscale}

The framework is not specific to any particular spacecraft class. Because the dynamics, allocation, and control layers operate on the abstract \texttt{Satellite} interface---mass, inertia tensor, and a list of actuators and sensors---the same code handles a 1U CubeSat and a large spacecraft. Defining a new configuration is a matter of specifying physical parameters:

\begin{lstlisting}
# 1U CubeSat (1.33 kg, MTQ-only)
sat_1u = ADCS.Satellite(mass=1.33,
    J_0=np.diag([0.002, 0.002, 0.002]),
    actuators=mtqs_small, sensors=sens)

# 3U CubeSat (4 kg, 3+1)
sat_3u = ADCS.satellite_factory.create_beavercube2_cubesat()

# 6U CubeSat (8 kg, 3+3)
sat_6u = ADCS.Satellite(mass=8,
    J_0=np.diag([0.08, 0.06, 0.06]),
    actuators=mtqs + rws_3axis, sensors=sens)

# Large satellite (500 kg, 3+3)
sat_large = ADCS.Satellite(mass=500,
    J_0=np.diag([50, 150, 150]),
    actuators=large_rws + large_mtqs, sensors=sens)
\end{lstlisting}

Each configuration plugs into the same simulation call, the same controller interface, and the same analysis and plotting tools. The lines of configuration code are approximately the same regardless of scale, as shown in Table~\ref{tab:multiscale}.

\begin{table}[h]
\centering
\caption{Framework generality across spacecraft scales (LP-PD controller, 1000\,s).}
\label{tab:multiscale}
\resizebox{\columnwidth}{!}{%
\begin{tabular}{lcccc}
\hline
Config & Mass (kg) & \# MTQ/RW & Config Lines$^*$ & Mean Error ($^\circ$) \\
\hline
1U  & 1.33 & 3/0 & $\sim$20 & 14.7 \\
3U  & 4 & 3/1 & $\sim$23 & 1.4 \\
6U  & 8 & 3/3 & $\sim$23 & 0.0 \\
Large  & 500 & 3/3 & $\sim$23 & 0.1 \\
\hline
\multicolumn{5}{l}{\footnotesize $^*$Including imports, initial state definition, simulate call, and } \\
\multicolumn{5}{l}{\footnotesize specification of goals, orbit, and controller.} \\
\end{tabular}%
}
\end{table}

The point is not that a 1U and a 500\,kg satellite achieve the same performance---they have very different actuation authority and disturbance environments. The point is that the same framework, the same control law, and the same workflow handles both without modification.

\subsection{Hardware-in-the-Loop on Raspberry Pi}\label{subsec:demo_hil}

The final demonstration moves the ADCS code from desktop simulation to flight-representative hardware. The same Python code used in the preceding demonstrations---identical source files, no recompilation or translation---runs on a Raspberry Pi~4, which serves as the flight computer for BeaverCube~2. It will run similarly on any computer that runs Python.

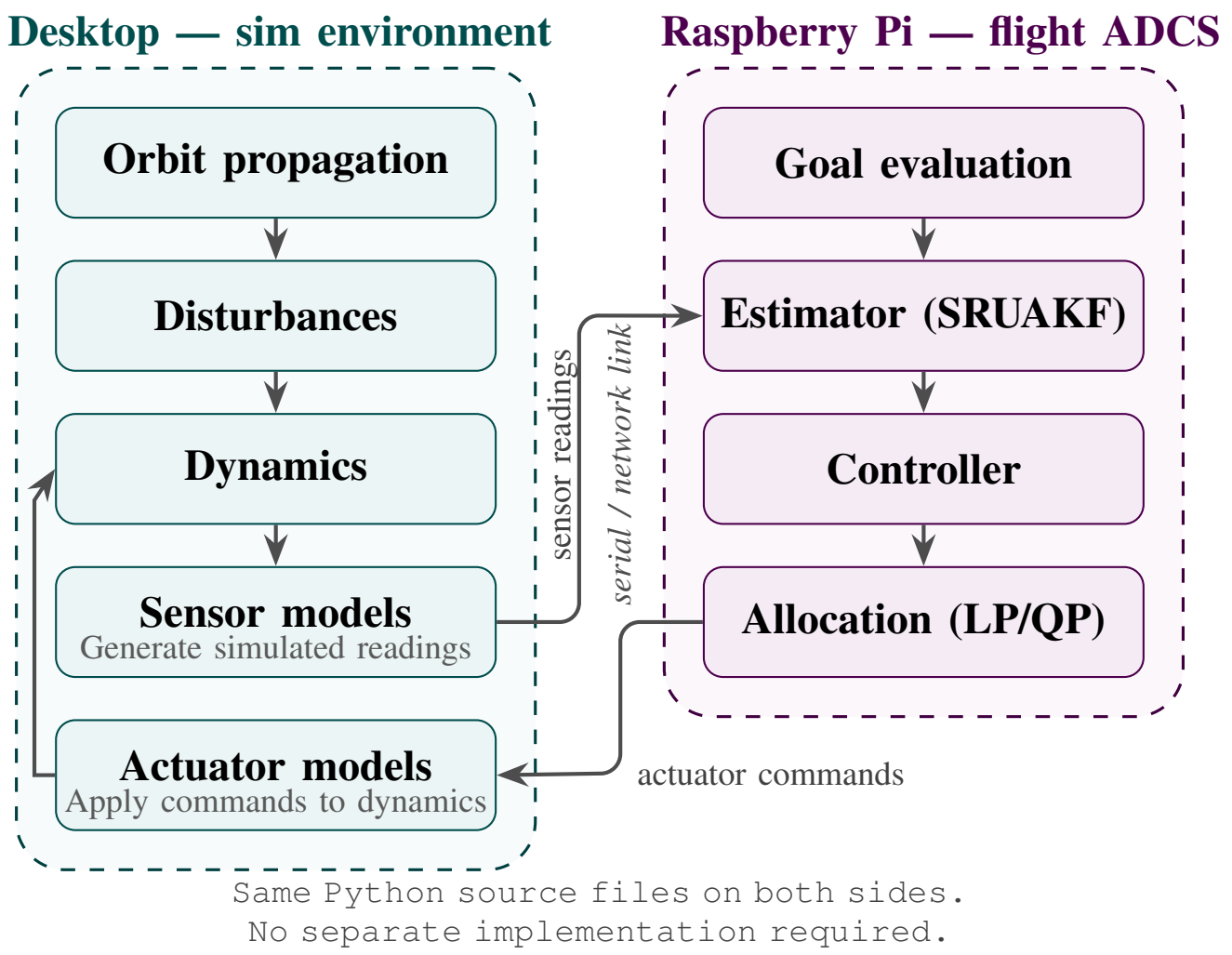
\begin{figure}[t]
\centering
\resizebox{\columnwidth}{!}{%
\begin{tikzpicture}[
    >=Stealth,
    node distance=0.4cm,
    box/.style={draw=gray!50, fill=gray!6, rounded corners=4pt,
                minimum height=0.8cm, line width=0.4pt, font=\small},
    simbox/.style={draw=teal!60!black, fill=teal!8, rounded corners=4pt,
                   minimum height=0.8cm, line width=0.4pt, font=\small,
                   minimum width=3.2cm},
    fswbox/.style={draw=violet!60!black, fill=violet!8, rounded corners=4pt,
                   minimum height=0.8cm, line width=0.4pt, font=\small,
                   minimum width=3.2cm},
    boxtitle/.style={font=\small\bfseries},
    boxsub/.style={font=\scriptsize, text=gray!70!black},
    flow/.style={->, gray!60!black, line width=0.8pt},
    flowlabel/.style={font=\scriptsize, text=gray!50!black},
    code/.style={font=\scriptsize\ttfamily, text=gray!40!black},
]

\node[simbox] (dynamics) at (0, 0) {};
\node[boxtitle] at (dynamics.center) {Dynamics};

\node[simbox, above=0.3cm of dynamics] (disturbances) {};
\node[boxtitle] at (disturbances.center) {Disturbances};

\node[simbox, above=0.3cm of disturbances] (orbit) {};
\node[boxtitle] at (orbit.center) {Orbit propagation};

\node[simbox, below=0.3cm of dynamics] (sensormod) {};
\node[boxtitle] at ([yshift=2pt]sensormod.center) {Sensor models};
\node[boxsub] at ([yshift=-6pt]sensormod.center) {Generate simulated readings};

\node[simbox, below=0.3cm of sensormod] (actmod) {};
\node[boxtitle] at ([yshift=2pt]actmod.center) {Actuator models};
\node[boxsub] at ([yshift=-6pt]actmod.center) {Apply commands to dynamics};

\begin{scope}[on background layer]
\node[draw=teal!50!black, dashed, line width=0.6pt, rounded corners=10pt,
      fill=teal!3, inner sep=8pt,
      fit=(orbit)(dynamics)(sensormod)(actmod)(disturbances)] (desktop) {};
\end{scope}
\node[font=\small\bfseries, text=teal!60!black, anchor=south west]
    at ([xshift=-5pt, yshift=-1pt]desktop.north west) {Desktop --- sim environment};

\node[fswbox, right=1.5cm of orbit] (goaleval) {};
\node[boxtitle] at (goaleval.center) {Goal evaluation};

\node[fswbox, below=0.3cm of goaleval] (estimator) {};
\node[boxtitle] at (estimator.center) {Estimator (SRUAKF)};

\node[fswbox, below=0.3cm of estimator] (controller) {};
\node[boxtitle] at (controller.center) {Controller};

\node[fswbox, below=0.3cm of controller] (allocation) {};
\node[boxtitle] at (allocation.center) {Allocation (LP/QP)};

\begin{scope}[on background layer]
\node[draw=violet!50!black, dashed, line width=0.6pt, rounded corners=10pt,
      fill=violet!3, inner sep=8pt,
      fit=(goaleval)(estimator)(controller)(allocation)] (rpi) {};
\end{scope}
\node[font=\small\bfseries, text=violet!60!black, anchor=south west]
    at ([xshift=-4pt, yshift=-1pt]rpi.north west) {Raspberry Pi --- flight ADCS};

\draw[flow, rounded corners=4pt]
    (sensormod.east) -- ++(0.6, 0) coordinate (mid-sens) -- ++(0, 2.23)
    -- (estimator.west);
\node[flowlabel, rotate=90, anchor=south,yshift=1pt,xshift=-5pt] at ($(desktop.east)!0.5!(rpi.west)$) {sensor readings};

\draw[flow, rounded corners=4pt]
    (allocation.west) -- ++(-0.6, 0) coordinate (act-top) -- ++(0, -1.1) coordinate (act-bot)
    -- (actmod.east);
\node[flowlabel, anchor=west] at (act-bot) {actuator commands};

\node[font=\scriptsize\itshape, text=gray!60!black, rotate=90, anchor=south,yshift=-10pt,xshift=-7pt]
    at ($(desktop.east)!0.5!(rpi.west)$) {serial / network link};

\draw[flow] (orbit) -- (disturbances);
\draw[flow] (disturbances) -- (dynamics);
\draw[flow] (dynamics) -- (sensormod);
\draw[flow] (actmod.west) -- ++(-0.15, 0) --  ++(0, 2) -- (dynamics.west);

\draw[flow] (goaleval) -- (estimator);
\draw[flow] (estimator) -- (controller);
\draw[flow] (controller) -- (allocation);

\node[font=\scriptsize\ttfamily, text=gray!50!black, anchor=north, text width=7cm, align=center]
    at ([yshift=-0.5cm]$(desktop.south)!0.5!(rpi.south)$)
    {Same Python source files on both sides.\\No separate implementation required.};

\end{tikzpicture}%
}
\caption{Hardware-in-the-loop architecture. The desktop simulates the physical environment; the Raspberry Pi runs the flight ADCS stack. Both use the same framework code---there is no translation step between simulation and flight software.}
\label{fig:hil}
\end{figure}

In the HIL configuration, a desktop computer runs the simulation environment: orbit propagation, disturbance torques, sensor signal generation, and actuator effect modeling. The Raspberry Pi runs the flight ADCS stack: attitude estimation (SRUAKF), goal management, control law execution, and torque allocation. The two communicate over a serial link, with the desktop sending simulated sensor readings and the Pi returning actuator commands, as depicted in Figure~\ref{fig:hil}.

\begin{table}
\centering
\caption{Raspberry Pi 4 ADCS loop timing.}
\label{tab:hil_timing}
\begin{tabular}{lc}
\hline
Metric & Value \\
\hline
Single control loop (ctrl + alloc) & 10 ms \\
Real-time factor (1\,s timestep) & 100 \\
Peak memory usage & 261 MB$^*$ \\
\hline
\multicolumn{2}{l}{\footnotesize $^*$Including all libraries, i.e. Skyfield.} \\
\end{tabular}

\end{table}

The timing numbers, shown in Table~\ref{tab:hil_timing}, confirm that Python is fast enough for CubeSat-class control loops, but the more important point is what code is being timed. This is the flight code---the same source files, unmodified, that could be executed on orbit. For a Python-based CubeSat stack, there is no separate flight implementation to validate against the simulation, no translation step where bugs can be introduced, and no question of whether the simulation accurately represents the flight software. It does, because it is the flight software. The HIL infrastructure itself is part of the open-source package: any team with a Raspberry Pi (or similar single-board computer) and a serial cable can run the same test, drastically simplifying HIL ADCS tests for small satellite teams and making them much more feasible.

\section{Comparison with Alternatives}\label{sec:comparison}

To substantiate the claim that this framework offers a simpler path to ADCS development, we implement an identical scenario---3U CubeSat, 3 MTQ + 1 RW, nadir pointing, ISS orbit (408\,km, 51.6$^\circ$), one orbit---in both our framework and the best available alternative, Basilisk. The results are shown in Table~\ref{tab:basilisk_head_to_head}. Both simulations use perfect state knowledge (no estimator) and a WMM magnetic field model.

\begin{table}[h]
\centering
\caption{Head-to-head comparison on identical scenario (3U, 3+1, nadir, 1 orbit, including GG/SRP/drag disturbances).}
\label{tab:basilisk_head_to_head}
\begin{tabular}{lcc}
\toprule
Metric & This Work & Basilisk \\
\midrule
Configuration lines & 45 & 213 \\
Unique classes required & 15 & 27 \\
Module instantiations & 20 & 35+ \\
Explicit message connections & 0 & 30 \\
Simulation runtime (5400\,s sim) & $\sim$38\,s & $\sim$0.4\,s \\\midrule
Lines to add estimation & $\sim$20 & N/A$^*$ \\
\bottomrule
\multicolumn{3}{l}{\footnotesize $^*$Full attitude estimation from TAM+CSS+gyro requires} \\
\multicolumn{3}{l}{\footnotesize custom C++ module or multi-filter composition.} \\
\end{tabular}
\end{table}

Basilisk is approximately 100$\times$ faster, owing to its compiled C++ dynamics core. This matters for Monte Carlo campaigns with thousands of runs (though our Monte Carlo configuration does allow parallel test cases, reducing the time difference). For the iterative single-run cycle that dominates early ADCS design---change configuration, run, inspect, adjust, repeat---both frameworks complete in acceptable time, and the 5$\times$ reduction in configuration effort compounds with every iteration.

The configuration difference reflects a deeper architectural choice. Basilisk's message-passing design offers flexibility for multi-domain simulation: each module is independently testable, and the explicit wiring makes data flow visible in the code. The cost is complexity for ADCS-focused tasks. Commanding three magnetic torque bars, for example, requires instantiating and wiring five modules in sequence plus constructing a configuration message payload; in this framework, the equivalent is one \texttt{ADCS.MTQ(axis, max\_torque)} call per torque bar. Changing a pointing objective from inertial hold to nadir requires replacing a guidance module and providing additional input messages; in this framework, it is \texttt{goal = ADCS.goals.Nadir\_Goal()}. These are not deficiencies in Basilisk's design---they are consequences of a general-purpose architecture applied to a specific task.

Where the difference is more fundamental is in estimation. In this framework, adding an SRUAKF estimator requires approximately 20 additional lines; the estimator automatically builds its measurement model from whatever sensors are present on the spacecraft. In Basilisk, full attitude estimation from a magnetometer, coarse sun sensor, and gyroscope---the sensor suite used by BeaverCube~2 and many CubeSats---is not available as a single module. The built-in estimators are sensor-specific (\texttt{inertialUKF} requires a star tracker; \texttt{sunlineSuKF} estimates only sun direction from CSS). Full attitude estimation from a TAM+CSS+gyro suite would require composing partial filters via \texttt{navAggregate} or writing a custom module in C++. This framework's estimators build their measurement models from the satellite object directly, handling arbitrary sensor suites without additional development.

Other tools occupy different niches. NASA's 42~\cite{stoneking_42} is a comprehensive multi-body simulator but its C codebase is not easily extensible for custom ADCS algorithms. GMAT~\cite{gmat} is orbit-focused and lacks closed-loop ADCS depth. The most common alternative is custom code---the default for most small satellite teams---which this framework is specifically designed to augment or replace: months of per-mission implementation reduced to configuration.

Basilisk is the more mature and capable tool. Its C++ core is essential for large simulation campaigns, its module library covers domains well beyond ADCS, and it has a larger developer community. This framework is not intended to replace it. The positioning is narrower: for teams whose primary concern is ADCS design and evaluation, a pure-Python tool with a simpler interface enables faster iteration when the bottleneck is developer time, not CPU time.

\section{Discussion}\label{sec:discussion}
\subsection{Better Testing Produces Better Results}\label{subsec:better_testing}

The demonstrations in Section~\ref{sec:demos} show two kinds of value that come from cheap iteration. The realism toggle (Section~\ref{subsec:demo_realism}) reveals how much performance changes as simulation fidelity increases: a controller that looks excellent under perfect-knowledge conditions may degrade significantly once estimation errors, sensor noise, and environmental disturbances are introduced. Running this comparison takes a few lines of configuration change, so there is no reason not to quantify the gap before committing to a design. The architecture trade (Section~\ref{subsec:demo_arch_trade}) illustrates a related point: ADCS capability does not scale linearly with hardware. Adding a single reaction wheel to a magnetorquer-only system reduces mean pointing error from 17.4$^\circ$ to 1.0$^\circ$---not a marginal improvement, but an entirely different class of performance. Discovering where these thresholds lie requires running the comparison, which is a three-line configuration change.

The result is that teams can afford to test more configurations and more fidelity levels and make better-informed design decisions. This framework makes those tests cheap enough to actually run. The automatic allocation layer is central to this: because it provides reasonable (though sub-optimal) performance on any combination of actuators, goals, and feedback laws without configuration-specific tuning, teams can evaluate ten architectures in an afternoon instead of spending a week tuning one.

\subsection{Flight Heritage}\label{subsec:flight_heritage}

The framework underpins the flight ADCS software for BeaverCube~2, a 3U Earth-observation CubeSat developed by MIT's STAR Lab in collaboration with Northrop Grumman. BeaverCube~2 carries visible and long-wave infrared imagers for ocean surface and cloud-top observation, with an onboard AI accelerator for feature identification. The ADCS consists of an ISISpace magnetorquer board and a single CubeSpace CubeWheel reaction wheel (the 3+1 configuration modeled throughout this paper), with no star trackers or propulsion. The mission was selected by NASA's CubeSat Launch Initiative and is currently in pre-launch integration.

The same Python code used in desktop simulation, Monte Carlo analysis, and hardware-in-the-loop testing on a Raspberry Pi will execute as the flight ADCS software. There is no separate flight implementation. This eliminates a class of bugs that arise from translating validated algorithms into a different language or framework for flight, and it means that every simulation run is testing the actual flight code. BeaverCube~1, a predecessor 3U mission with a 3+0 (magnetorquer-only) ADCS, launched in 2022 aboard SpaceX CRS-25 and was deployed from the International Space Station; an earlier version of this framework was used for its ADCS development and testing as well.\footnote{Sadly, BeaverCube-1 suffered a radio issue and was never heard from, so we do not have data on its ADCS performance.}

\subsection{Student Team Adoption}\label{subsec:student_teams}
The framework is being provided to university satellite teams to evaluate its accessibility. In structured one-hour introductions, teams with no prior ADCS development experience go from installation to running and modifying simulations with their own spacecraft parameters, in only an hour or two.

The real test of accessibility is not whether the developers can use their own tool, but whether new teams---with different spacecraft, different missions, and varying levels of experience---can get useful results without extensive support. Early adoption experience suggests that the barrier to entry is low enough for ADCS development to fit within a single design sprint rather than consuming months of dedicated effort.

\subsection{Limitations}\label{subsec:limitations}

The framework has several known limitations. Python execution speed is slower than C++ or Fortran alternatives; while acceptable for CubeSat-class control loops (sub-100\,ms on a Raspberry Pi~4), it is not suitable for high-rate applications or very long Monte Carlo campaigns without patience.

The framework's scope is deliberately limited to ADCS. It does not model power, thermal, communications, or structural subsystems, and its environmental models---while adequate for attitude control design---are not intended for precision orbit determination. Users requiring multi-physics fidelity should consider coupling this framework's ADCS output with a more comprehensive mission simulator.

Finally, this project is not complete--it is the sort of framework that can always be expanded. For example, the current estimation subsystem does not include inertia estimation. For missions where mass properties change over the mission lifetime (e.g., due to deployables or propellant consumption), the estimator cannot currently track these changes. Adding inertia as an augmented state is a planned extension. Similarly, we hope to include support for other actuators and sensors, different noise models, new control laws and estimator formulations, and more as time continues.

\section{Conclusion}\label{sec:conclusion}

This paper presented an open-source, modular ADCS framework in pure Python that makes attitude control development accessible to small satellite teams. From a single \texttt{pip install}, users reach a working simulation in under five minutes; from there, architecture trades, control law comparisons, fidelity sweeps, and hardware-in-the-loop testing are configuration changes rather than reimplementation efforts. Demonstrations show that a three-line actuator change reveals order-of-magnitude pointing differences between configurations, and that the same unmodified source code runs on a laptop, in Monte Carlo campaigns, and as flight software on a Raspberry Pi. The framework underpins the flight ADCS for BeaverCube~2, providing credibility beyond simulation-only validation. A head-to-head comparison with Basilisk shows a tradeoff of 100$\times$ slower runtime for 5$\times$ fewer configuration lines---a tradeoff that favors rapid iteration when the bottleneck is developer time, not CPU time. The framework is available at \url{https://github.com/nscheuer/Generalized_ADCS}.

\section*{Acknowledgments}
The authors would like to thank Alex Rose (MIT) for their contributions in developing the early version of this framework. Additionally, the authors would like to thank William Goldman (Tufts University) and Alexander Pati\~{n}o Walb{\ae}kken (ETH Z\"{u}rich) for helping develop the package.

\bibliographystyle{IEEEtran}
\bibliography{my_refs}

\end{document}